\documentclass[a4paper,11pt]{article}
\usepackage{pos}
\usepackage{subfig}
\usepackage{amsmath}
\usepackage{graphicx}
\usepackage{xspace}
\usepackage{lineno}
\title{Searches for leptoquarks in scenarios of lepton flavor universality anomalies}

\manuallySeparateAuthors
\author*{Arne Christoph Reimers}
\author{on behalf of the ATLAS and CMS Collaborations}

\affiliation{Universit\"at Z\"urich, Switzerland}

\emailAdd{arne.reimers@cern.ch}

\abstract{Leptoquarks with masses at the TeV scale have been proposed as possible solutions to flavor anomalies reported in the b-flavor sector. Based on data taken in proton-proton collisions at $\sqrt{s} = 13\,\mathrm{TeV}$ at the LHC, different leptoquark flavors can be probed in various final states. In this article, the results of direct searches for single and pair production of leptoquarks conducted by the ATLAS and CMS Collaborations are summarized.}

\FullConference{%
  *** Particles and Nuclei International Conference - PANIC2021 ***\\
  *** 5 - 10 September, 2021 ***\\
  *** Online ***
}

\newcommand{\tev}{\ensuremath{ \,\text{TeV} }\xspace}
\newcommand{\fbinv}{\ensuremath{{\,\text{fb}}^{-1}}\xspace}

\newcommand{\LQ}{\ensuremath{\mathrm{LQ}}\xspace}

\newcommand{\mlq}{\ensuremath{M_{\LQ}}\xspace}
\newcommand{\B}{\ensuremath{\mathcal{B}}\xspace}

\begin{document}
\maketitle

\section{Introduction}

Measurements of anomalous B meson decays, in particular via $\mathrm{b}\to\mathrm{c}\ell\nu$~\cite{HFLAV2021} and $\mathrm{b}\to\mathrm{s}\ell\ell$~\cite{LHCb:2021trn} transitions, hint at the violation of lepton flavor universality (LFU) in nature. If confirmed in future measurements, LFU violation would unambiguously imply the presence of physics beyond the standard model (SM). The existence of TeV-scale leptoquarks (LQs), new hypothetical scalar or vector particles mediating interactions between quarks and leptons, is one of the explanations most widely discussed in the literature~\cite{Cornella:2021sby}.
In general, the production and decay of LQs depend on three parameters: the LQ mass \mlq, the flavor-dependent LQ-quark-lepton coupling $\lambda$, and the branching fraction $\mathcal{B}$ of the LQ decay. For $\B=0$, the LQ couples only to quarks and neutrinos, while for $\B=1$ it only decays to a quark and a charged lepton.

\section{Leptoquark searches at ATLAS and CMS}

LHC searches for LQs have previously been performed by the ATLAS~\cite{ATLAS:2008xda} and CMS~\cite{Chatrchyan:2008aa} Collaborations at $\sqrt{s}=8\tev$. Here, an overview of the status of LQ searches at $\sqrt{s}=13\tev$ is given. The analyses presented have been performed on datasets corresponding to between 35.9 and 139\fbinv. No deviation from the SM expectation is observed in any analysis and lower limits on the LQ mass are placed.

\subsection{Leptoquarks decaying to $\mathrm{q}\ell$ or $\mathrm{b}\ell$}
\begin{figure}[t]
  \centering
  \resizebox{.99\textwidth}{!}{
    \includegraphics[height=5cm]{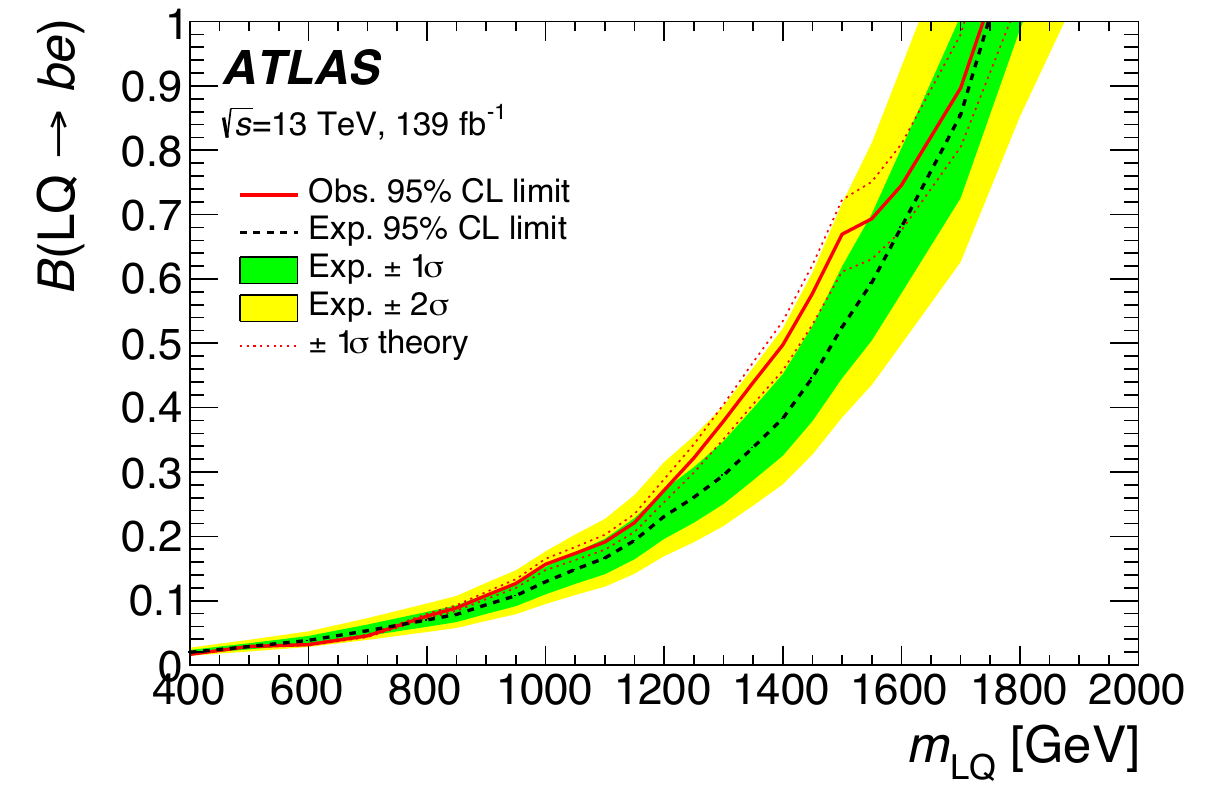}
    \quad
    \includegraphics[height=5cm]{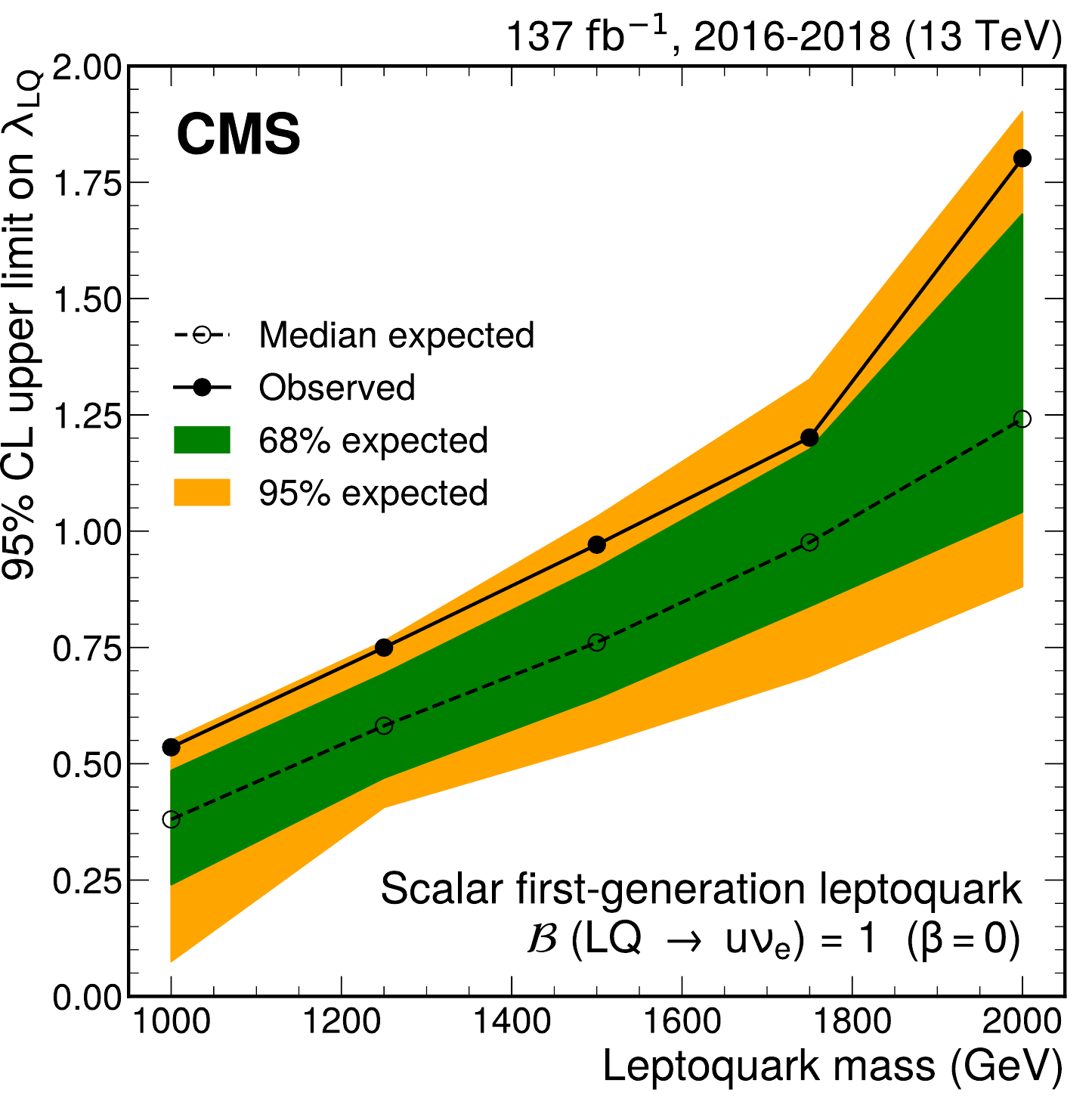}
  }
  \caption{Mass exclusion limits for LQs decaying to bottom quarks and electrons for varying \B~\cite{ATLAS:2020dsk} (left) and LQs coupled to up quarks and neutrinos for varying $\lambda$~\cite{CMS:2021far} (right).}
  \label{fig:blqnu}
\end{figure}

The ATLAS Collaboration has performed a search for pair-produced LQs coupled to $\mathrm{q}\ell$ or $\mathrm{b}\ell$~\cite{ATLAS:2020dsk}, where $\ell$ denotes an electron or a muon and q refers to an up, down, charm, or strange quark. Data corresponding to 139\fbinv have been analyzed. Event categories are defined using b- and c-tagged jets, allowing to specifically target LQ decays to $\mathrm{c}\ell$ and $\mathrm{b}\ell$ for the first time. Lower limits on the mass of scalar LQs of between $1.7$ and $1.8\tev$ are set under the assumption of $\B=1$ for all quark and lepton flavors considered.
As illustrated in Figure~\ref{fig:blqnu} (left) for $\LQ\to\mathrm{be}$ decays, the limits extend to lower values of \B, excluding scalar LQs below masses of about $0.8\tev$ for $\B=0.1$.

The ATLAS search for an asymmetry in $e^+\mu^-$ and $e^-\mu^+$ production using 139\fbinv of data~\cite{ATLAS-CONF-2021-045} has been interpreted in a scalar LQ scenario. The presence of such an LQ coupled to first- and second-generation fermions would enhance $e^+\mu^-$ production via single LQ production. Lower limits on the scalar LQ mass are placed as a function of $\lambda$.
Assuming $\B=0$ and $\lambda=1$, LQs below masses of approximately 1.75\tev are excluded.

\subsection{Leptoquarks decaying to $\mathrm{q}\nu$, $\mathrm{b}\nu$, or $\mathrm{t}\nu$}
\label{subsec:b0}
The ATLAS and CMS results on LQs coupled exclusively to quarks and neutrinos are reinterpretations of searches for squark pair production with decays to quarks and neutralinos, assuming vanishing neutralino masses. The CMS Collaboration has performed an analysis of $35.9\fbinv$ of data and excludes scalar LQs below masses of about $1.0\tev$~\cite{CMS:2018qqq} for all quark flavors assuming $\B=0$, while vector LQs are excluded up to higher masses due to their larger production cross section. Using data corresponding to 139\fbinv, the ATLAS Collaboration excludes scalar LQs decaying exclusively to $\mathrm{b}\nu$ or $\mathrm{t}\nu$ up to masses of about 1.25\tev for $\B=0$~\cite{ATLAS:2021yij, ATLAS:2020dsf}. Here, limits are also set for varying values of \B for both couplings.

The CMS mono-jet search~\cite{CMS:2021far} has been interpreted in a scenario of singly and pair-produced LQs coupling to $\mathrm{u}\nu$. Limits are set in the \mlq--$\lambda$--plane, as shown in Figure~\ref{fig:blqnu} (right). For $\lambda=1$, scalar LQs below masses of about 1.5\tev are excluded under the assumption of $\B=0$.

\subsection{Leptoquarks decaying to either $\mathrm{b}\tau$ or $\mathrm{t}\nu$}

\begin{figure}[t]
  \centering
  \resizebox{.99\textwidth}{!}{%
    \includegraphics[height=5cm]{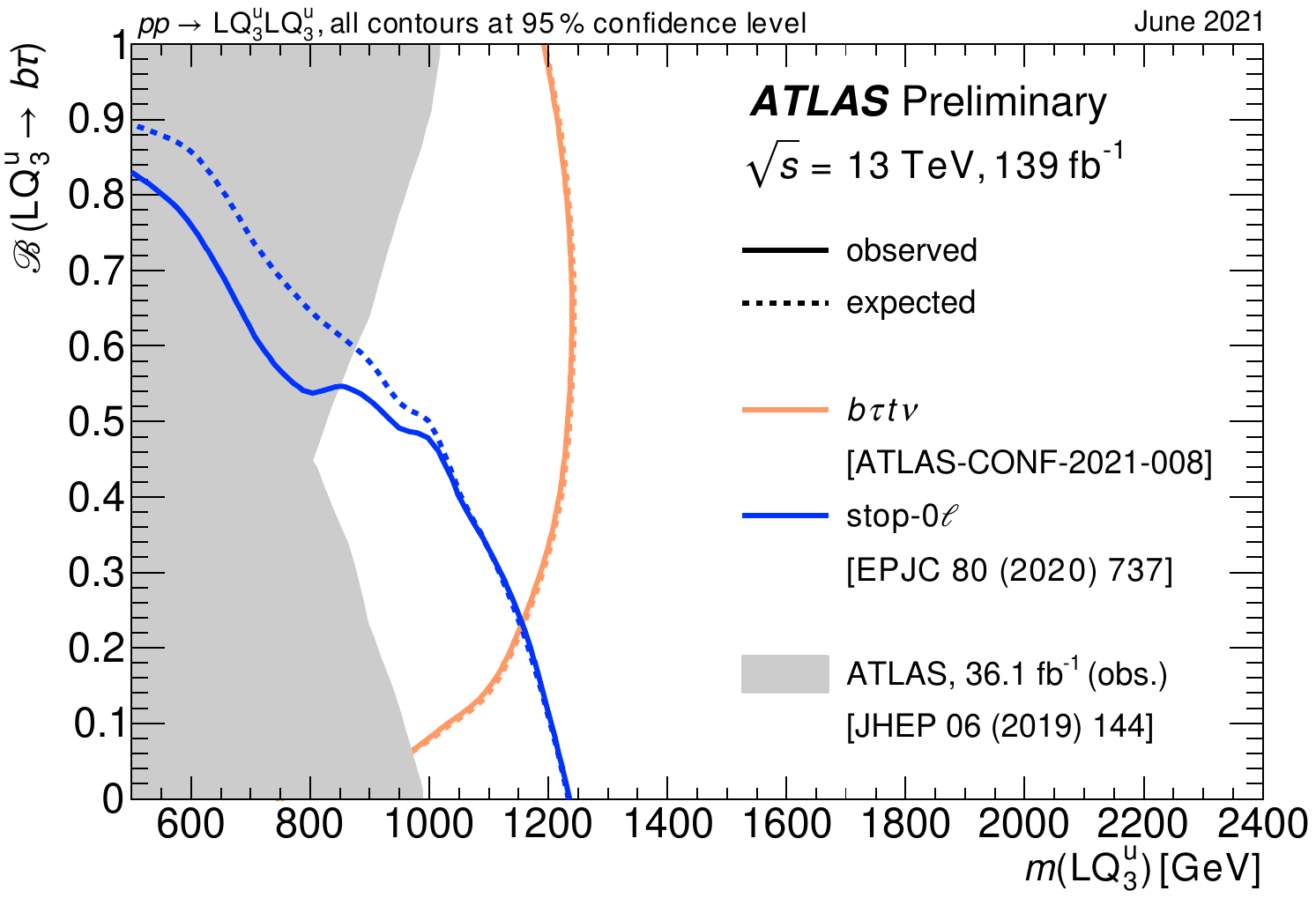}%
    \quad
    \includegraphics[height=5cm]{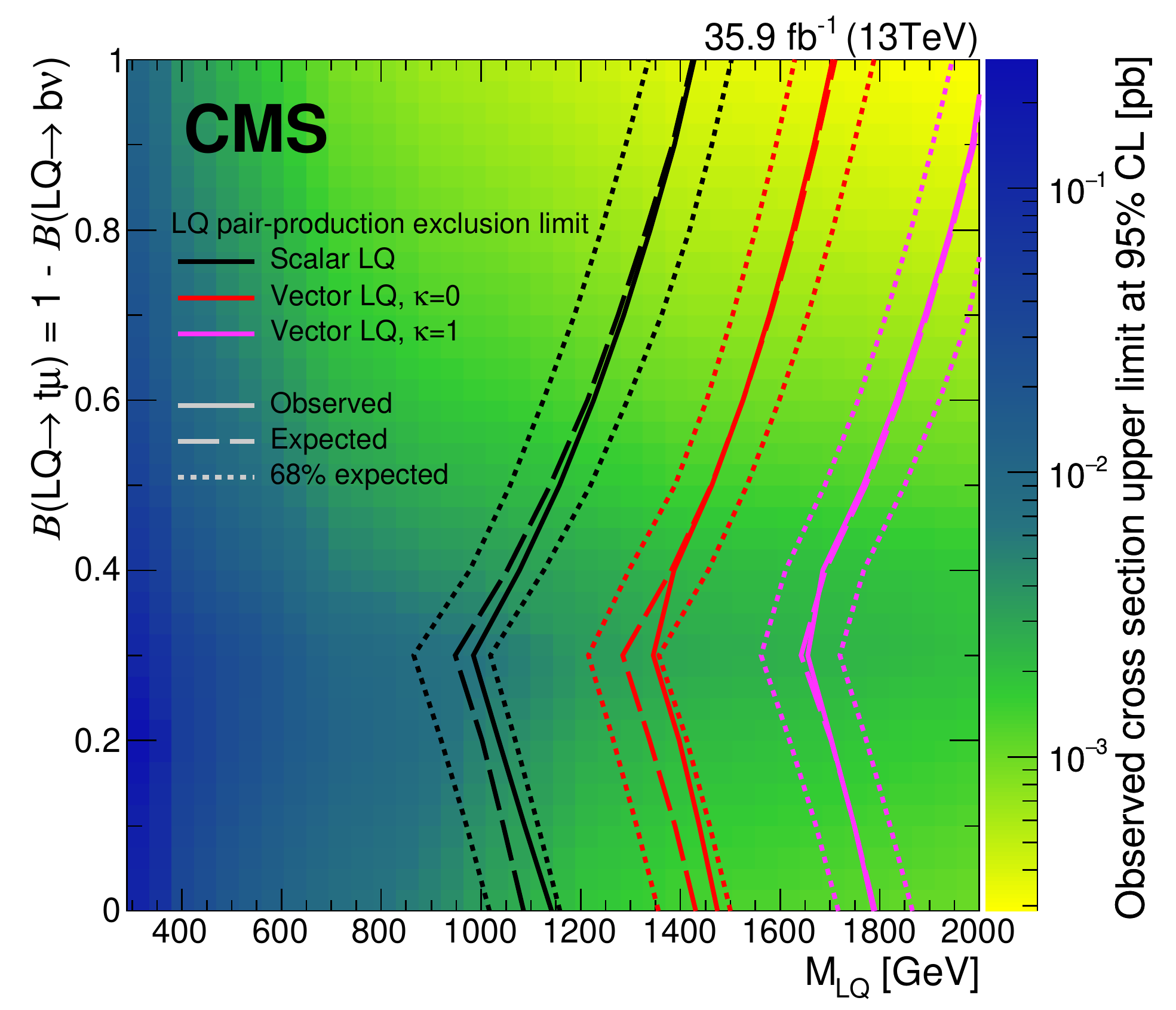}%
  }
  \caption{Overlay of exclusion limits as a function of \B for LQs decaying to either $\mathrm{b}\tau$ or $\mathrm{t}\nu$~\cite{ATL-PHYS-PUB-2021-017} (left) or LQs decaying to either $\mathrm{t}\mu$ or $\mathrm{b}\nu$~\cite{CMS:2018oaj} (right).}
  \label{fig:btautmu}
\end{figure}

Leptoquarks decaying to a bottom quark and a $\tau$ lepton are suited particularly well to explain the $\mathrm{b}\to\mathrm{c}\ell\nu$ anomalies.
CMS analyses, focusing on single~\cite{CMS:2018txo} and pair~\cite{CMS:2018iye} production using 35.9\fbinv of data, exclude scalar LQs with masses below $1.0\tev$ assuming $\B=1$.
Since the same LQs could also decay to $\mathrm{t}\nu$, the scenario corresponding to $\B=0.5$ has been studied by both collaborations using the full $\sqrt{s}=13\tev$ dataset. The ATLAS result~\cite{ATLAS:2021jyv} sets a lower mass limit of 1.25\tev (1.8\tev) on scalar (vector) LQs, while the CMS Collaboration has excluded vector LQs for $\mlq\lesssim1.65\tev$~\cite{CMS:2020wzx} combining the single and pair production modes.
Considering the results for $\B=0$ presented in section~\ref{subsec:b0}, all values of $\B$ have been probed in LHC searches, which is illustrated by the overlay of ATLAS limits shown in Figure~\ref{fig:btautmu} (left).

\subsection{Leptoquarks decaying to either $\mathrm{t}\ell$ or $\mathrm{b}\nu$}
The ATLAS Collaboration has conducted a search for pair-produced scalar LQs that decay to $\mathrm{te}$ or $\mathrm{t}\mu$~\cite{ATLAS:2020xov} using 139\fbinv of data.
At $\B=1$, values of $\mlq\lesssim 1.5\tev$ are excluded for both lepton flavors, probing the $\LQ\to\mathrm{te}$ decay for the first time. At $\B\simeq 0.15$, the lower limit on the LQ mass reaches 0.9\tev.
The CMS result for LQ pair production in the $\LQ\to\mathrm{t}\mu$ decay channel has been obtained using a dataset corresponding to 35.9\fbinv~\cite{CMS:2018oaj}. A lower limit on scalar LQ masses of 1.4\tev is placed assuming $\B=1$. Considered together with the results in the $\mathrm{b}\nu$ channel, scalar LQs are excluded below 0.9\tev for any value of $\B$, as shown in Figure~\ref{fig:btautmu} (right). More stringent limits of up to 2.0\tev are placed on vector LQs.

\subsection{Leptoquarks decaying to either $\mathrm{t}\tau$ or $\mathrm{b}\nu$}

\begin{figure}[tb]
  \centering
  \resizebox{.99\textwidth}{!}{%
    \includegraphics[height=5cm]{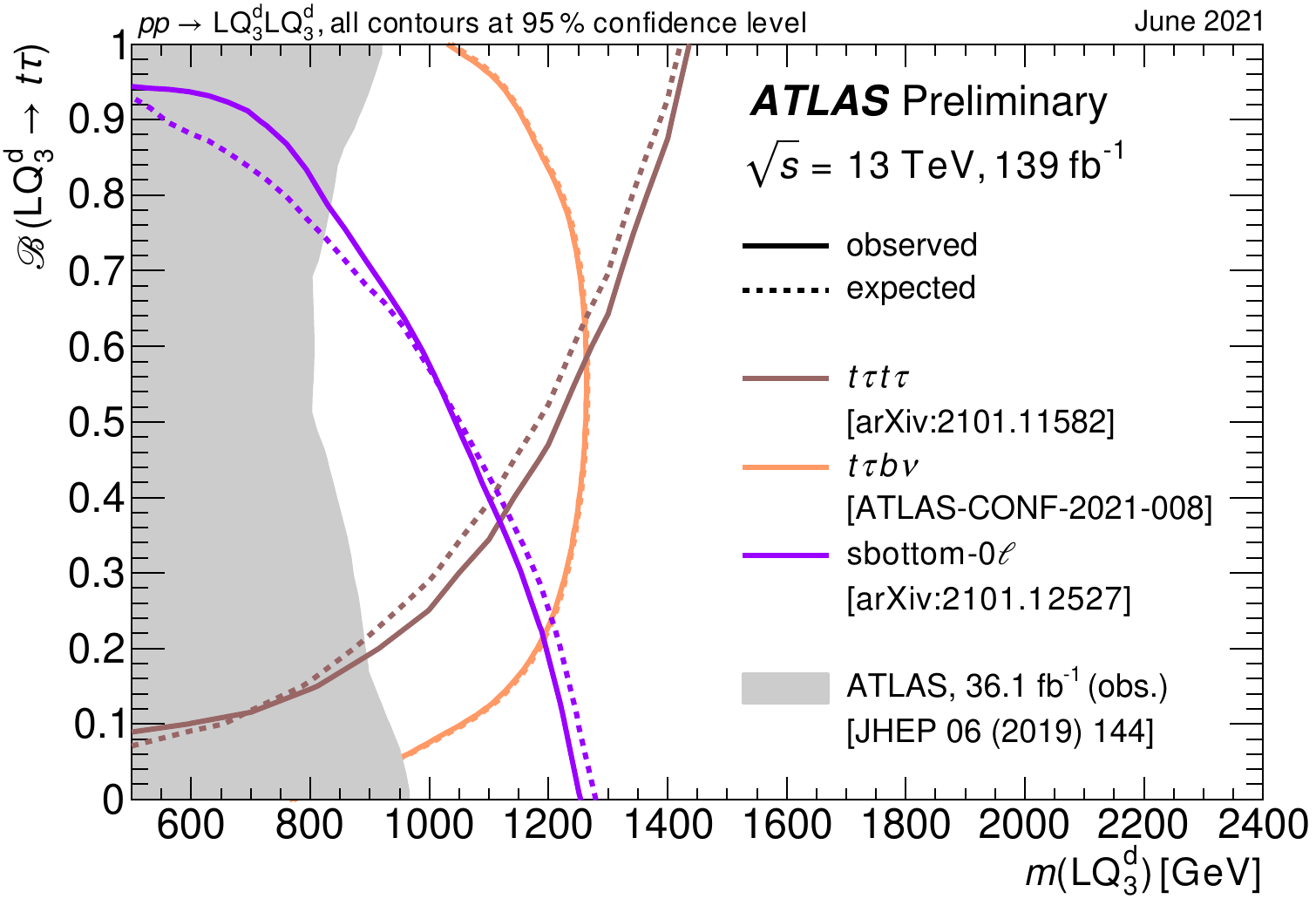}%
    \quad
    \includegraphics[height=5cm]{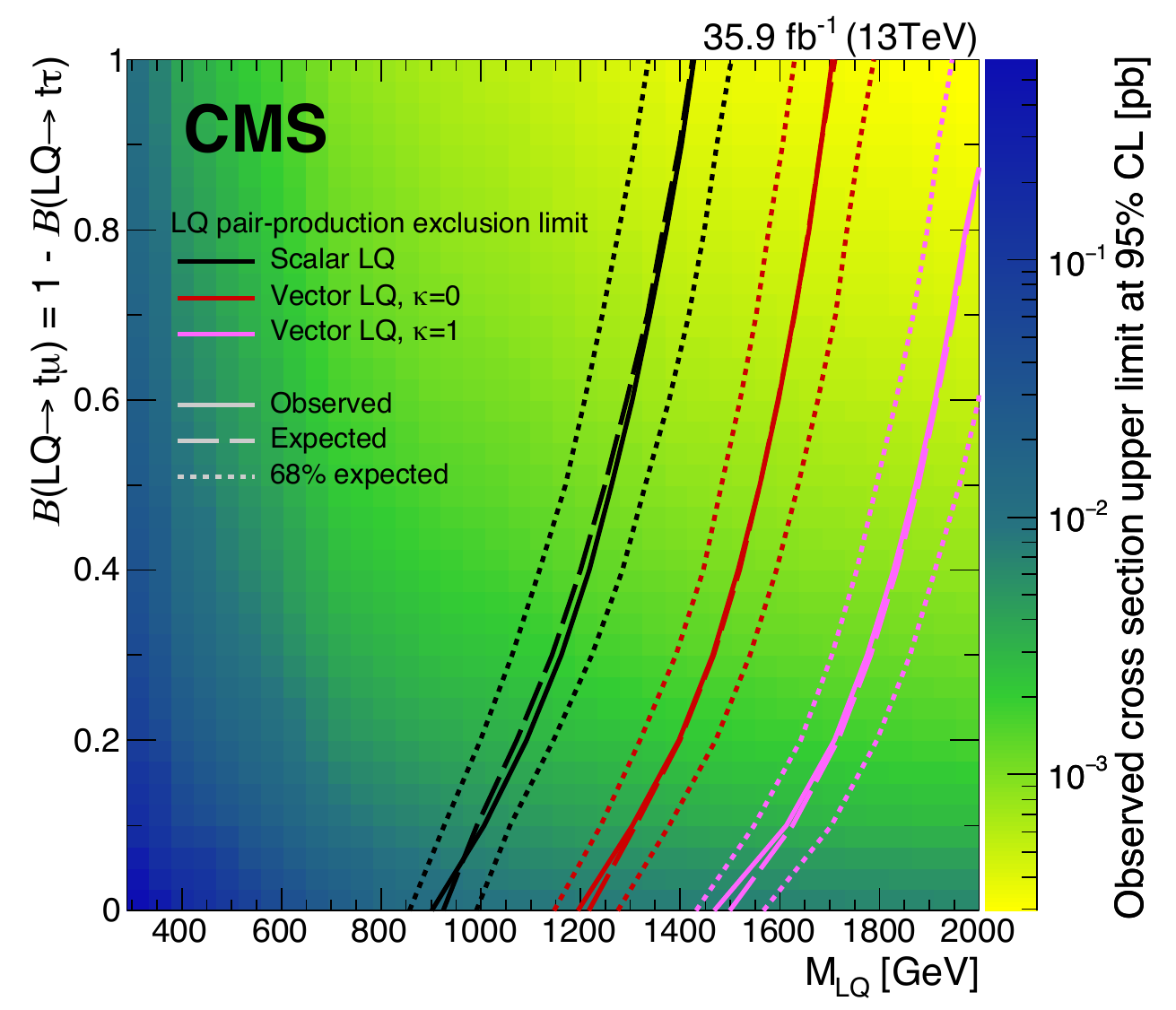}%
  }
  \caption{Overlay of exclusion limits for LQs decaying to either $\mathrm{t}\tau$ or $\mathrm{b}\nu$ as a function of \B~\cite{ATL-PHYS-PUB-2021-017} (left) or LQs decaying to either $\mathrm{t}\tau$ or $\mathrm{t}\mu$ for varying branching fraction~\cite{CMS:2018oaj} (right).}
  \label{fig:ttau}
\end{figure}

Leptoquarks decaying to $\mathrm{t}\tau$ or $\mathrm{b}\nu$ have been sought for systematically by the ATLAS and CMS Collaborations.
In searches that target exclusive decays to $\mathrm{t}\tau$ ($\B=1$) using 139 and 35.9\fbinv of data, ATLAS~\cite{ATLAS:2021oiz} and CMS~\cite{CMS:2018svy} exclude pair-produced scalar LQs below masses of 0.9 and about 1.4\tev, respectively.
The case of $\B=0.5$ has been studied by both ATLAS and CMS using datasets of 139 and 137\fbinv, respectively. The CMS Collaboration excludes scalar LQs up to masses of approximately 0.95\tev in a combined analysis of pair and single LQ production~\cite{CMS:2020wzx}, while the ATLAS Collaboration focuses on pair-produced scalar LQ and excludes $\mlq\lesssim1.25\tev$~\cite{ATLAS:2021jyv}.
In Figure~\ref{fig:ttau} (left), the exclusion limits placed by the ATLAS Collaboration are overlaid as a function of \B and \mlq, including the result in the $\mathrm{b}\nu$ decay mode. Scalar LQs are excluded below masses of 1.2\tev for all values of \B. The CMS Collaboration has considered LQ decays with varying branching fraction for the $\mathrm{t}\tau$ and $\mathrm{t}\mu$ decay modes. The cross section upper limits and lower limits on the mass of scalar and vector LQs are obtained from a statistical combination of the two analyses focusing on these decay channels. The results are presented in Figure~\ref{fig:ttau} (right) as a function of the LQ mass and the branching fraction considered.

\section{Conclusion}
Leptoquarks have been proposed as an explanation for the anomalies measured in B meson decays. The ATLAS and CMS Collaborations have conducted a comprehensive program of searches for LQs in the 13\tev era of the LHC, recently including single LQ production modes. No significant deviation from the standard model expectation has been observed, setting lower limits on the mass of scalar and vector LQs consistently exceeding 1\tev.

\bibliographystyle{JHEP_short}
\bibliography{bibliography_short}

\end{document}